\documentclass[aps, reprint,showpacs,showkeys,pra,nofootinbib]{revtex4-1}
\usepackage{dcolumn}
\usepackage[utf8]{inputenc}
\usepackage{amsmath}
\usepackage{amsfonts}
\usepackage{amssymb}
\usepackage{graphicx}
\usepackage{hyperref}
\usepackage{xcolor}

\newcommand{\T}{\text}
\newcommand{\rme}{\mathrm{e}}

\begin{document}

\title{Effective electrostatic attraction between electrons due to quantum interference}

\author{Marina F. B. Cenni}\author{Raul Corr\^ea}\email{Present address: Faculty of Physics, University of
Warsaw, Pasteura 5, 02-093 Warsaw, Poland
}\author{Pablo L. Saldanha}\email{saldanha@fisica.ufmg.br}
\affiliation{Departamento de F\'isica, Universidade Federal de Minas Gerais, Caixa Postal 701, 30161-970, Belo Horizonte, MG, Brazil}

\date{\today}

\begin{abstract}
We show how, with the use of quantum interference, we can violate, in some sense, the rule that charges of equal sign always repel each other. By considering two electrons that propagate parallel to each other in a Mach-Zehnder interferometer, we show that the quantum superposition of the electrostatic repulsion when the electrons propagate in the same path with the absence of interaction when they propagate in opposite paths may result in an effective attraction between them, when we post-select by which port each electron leaves the interferometer. We also discuss an experimental setup that could be used to verify such an effect.
\end{abstract}



\maketitle

\section{Introduction}
Electric charges of equal sign repel each other, while charges of opposite sign experience an attraction. This sentence represents a fundamental principle not only for the scientific community but that is well known by the general public as well. It is hard to imagine that such principle could be violated. But here we show that this principle may be contradicted by a quantum interference phenomenon with post-selection. 

Quantum interference may result in many non-intuitive phenomena such as interferometry with massive particles \cite{feynman3,tonomura89,bach13}, quantum delayed choice experiments \cite{ionicioiu11,peruzzo12,kaiser12}, quantum erasers \cite{scully91,herzog95,durr98,walborn02}, ``interaction-free'' measurements \cite{elitzur93,kwiat95,peise15} or the Hong-Ou-Mandel effect \cite{hongoumandel87,bocquillon13,mahrlein16}. Here we discuss an example of the counter-intuitive characteristics of quantum interference inspired by a recent work by Aharonov \textit{et al.} \cite{aharonov132} and following the discussions of an earlier paper from our group \cite{qiof}. In Ref. \cite{aharonov132} the authors discuss the classical limit of the radiation pressure and the difference in interpretation arising from classical and quantum descriptions, by treating one of the mirrors of an interferometer quantum mechanically. The authors have shown how it is possible for the quantum combination of two possibilities, one in which light pushes a mirror outwards and other that leaves it still, to result in a inward pull in the mirror. Our previous paper generalizes this result by considering anomalous shifts in momentum associated with general quantum objects, and proposes feasible ways of testing the effect in the laboratory outside the weak interaction regime \cite{qiof}. With this we have introduced the concept of the ``quantum interference of force'' effect. 

Here we describe an interesting phenomenon based on the quantum interference of force effect \cite{qiof}, by considering two electrons that propagate parallel to each other in a Mach-Zehnder interferometer and post-selecting the interferometer port where each electron exits. In this two-electron interferometer, the electrons will interact or not based on the paths taken by each of them inside the system, resulting in entanglement between the electrons' states. A related analysis can be found in \cite{dressel12}, where the system consists of two single-electron Mach-Zehnder interferometers coupled by Coulomb interaction, and the post-selection of one electron exit is used to obtain path-information about the other electron due to the existence of the interaction phase. Here we show that the quantum superposition of the situations where the electrons propagate in the same interferometer arm, repelling each other, with the situations where they propagate in opposite arms, with no interaction, may result in an effective  attraction between them. This effective electrostatic attraction between the electrons is manifested in the momentum distribution of each electron, that changes its mean value in the direction of the other electron with the propagation through the interferometer.

\section{Setup description}
\begin{figure}
				\centering
					\includegraphics[width=7.1cm]{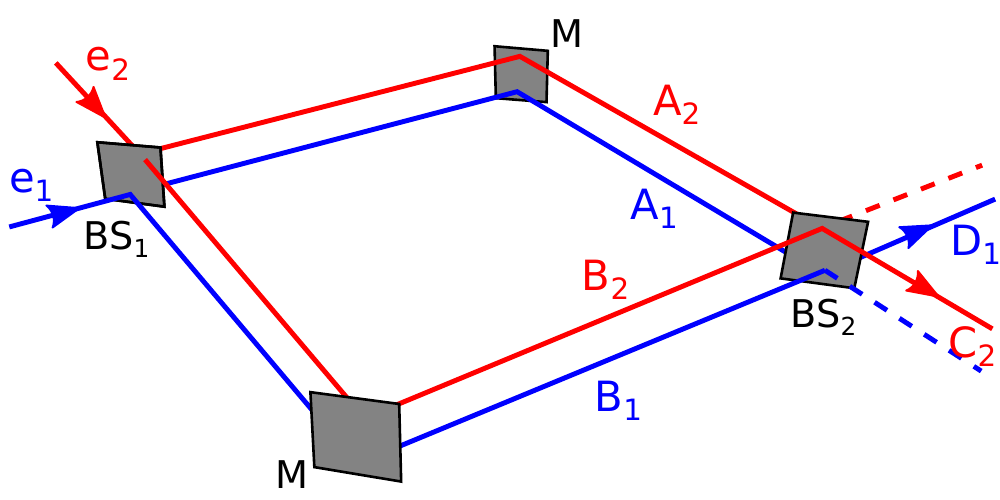}
				\caption{Two electrons propagate parallel to each other in a  Mach-Zehnder interferometer, entering by the indicated ports. Beam splitter BS$_1$ splits the incident wave functions and the mirrors M redirect the electrons to interfere at the second beam splitter BS$_{2}$. The lines represent the centers of the wave functions of the electrons e$_{1}$ (blue) and e$_{2}$ (red) while propagating in the interferometer. The distance $d$ between the electrons' paths is considered to be much larger than the widths of their wave functions, such that the electrons can be labeled as e$_{1}$ and e$_{2}$ due to their spatial distinguishability. If the electrons propagate in the same arm they repel each other, while if they propagate in opposite arms their interaction is negligible. We will post-select the events where electron e$_1$ exits by $D_1$ and electron e$_2$ exits by $C_2$.} 
				\label{fig28}
			\end{figure}

Consider a two-path Mach-Zehnder interferometer with two electrons e$_1$ and e$_2$ sent at the same time through the apparatus, as depicted in Fig. \ref{fig28}. The electrons can be distinguished from one another by the $x$ component of their position, with their separation $d$ being much larger than the width of their wave functions. Apart from this displacement, the states of the electrons are essentially the same. Both paths are considered to be free from any external influence and isolated from each other so that only the electromagnetic interactions between electrons taking the same path are allowed to take place. We associate the orthogonal state vectors $|A_i\rangle$ and $|B_i\rangle$ with the distinguishable paths of propagation possible for the electrons during their travel through the system, and the vectors $|C_i\rangle$ and $|D_i\rangle$ with the possible exit ports of the Mach-Zehnder interferometer, matching the labeling given by Fig. \ref{fig28}. The reflection and transmission coefficients for each beam splitter, BS$_{1}$ and BS$_{2}$, are the same, denoted by $ir$, and $t= \sqrt{1-r^2}$, with $r$ and $t$ real. The initial quantum state of the $x$ component of the electrons' momentum will be denoted by $|\Phi_i\rangle$, with $i=\{1,2\}$.
			
			  We consider a post-selection of the totality of events where electron e$_{1}$ exits the interferometer by {$D_{1}$} and e$_{2}$ exits by {$C_{2}$}, as indicated in Fig. \ref{fig28}. By considering this post-selection choice, the final joint state of the system that consists of the two electrons will be a coherent sum over the amplitudes associated with all the possible ways for this system to have evolved in time towards this final state. There are in total four possibilities of evolution for the described system: two where the electrons take different paths inside the interferometer and therefore do not interact, and two where they do travel by the same path and an electric interaction between them exists during some time interval. In the first two cases where the electrons do not interact, the state of the system will evolve as
	
	\begin{itemize}
		\item e$_{1}$ goes through path $A_{1}$ and e$_{2}$ goes through $B_{2}$:
\begin{equation}
-r^2t^2e^{i\phi}|\Phi_{1},D_{1}\rangle|\Phi_{2},C_{2}\rangle ,
\end{equation}			
\item e$_{1}$ goes through path $B_{1}$ and e$_{2}$ goes through $A_{2}$:
\begin{equation}
-t^2r^2e^{i\phi}|\Phi_{1},D_1\rangle|\Phi_{2},C_2\rangle ,
\end{equation}		
	\end{itemize}
where the vector states associated with each electron individually are labeled accordingly, and $\phi$ represents an extra phase for an electron propagation through path $A_i$ in relation to a propagation through path $B_i$.
				
				In turn, considering that the interaction between the electrons will change their momentum states, the quantum state associated with the last two possibilities of evolution where the electrons take the same path and therefore interact will evolve as
\begin{itemize}
	\item e$_{1}$ goes through path {$A_{1}$} and e$_{2}$ goes through {$A_{2}$}:
\begin{equation}\label{pa}
-r^2t^2e^{i(2\phi+\alpha)}|\Phi^{-}_{1},D_{1}\rangle|\Phi^{+}_{2},C_{2}\rangle ,
\end{equation}
\item e$_{1}$ goes through path {$B_{1}$} and e$_{2}$ goes through {$B_{2}$}:
\begin{equation}\label{pb}
-r^2t^2e^{i\alpha}|\Phi^{-}_{1},D_1\rangle|\Phi^{+}_{2},C_2\rangle,
\end{equation}
\end{itemize}
where we have taken the vectors $|\Phi^{\mp}_{i}\rangle$ to represent the electrons' momentum states that were disturbed by their electromagnetic interaction, and $\alpha$ represents a phase gained due to the interaction. Considering the combination of these four probability amplitudes, the post-selected electrons' momentum state is
\begin{equation}\label{entstate}
|\Phi_{ps}\rangle \propto |\Phi_{1}\rangle|\Phi_{2}\rangle + e^{i\alpha}\cos{(\phi)}|\Phi^{-}_{1}\rangle|\Phi^{+}_{2}\rangle.
\end{equation}

\section{Results}
       To closely analyze these results, we shall specify the initial wave functions for the $x$ component of the electrons' momentum $\Phi_i(p)=\langle p|\Phi_i\rangle$ as Gaussian distributions with width $W$ centered at zero:
\begin{equation}
\Phi_i(p)=\frac{\pi^{-\frac{1}{4}}}{\sqrt{W}}\ \text{exp}\left[-\frac{1}{2}\left(\frac{p}{W}\right)^2\right],\label{gauss}
\end{equation}
where the origin of the $x$ axis for each electron was defined at the corresponding center of its position wave function. If the electrons' separation is much larger than the width of their wave functions and if this width does not change appreciably during the electrons' time travel along the interferometer, the electrons' interaction results in  shifts $\delta$ on their momentum wave functions without altering their Gaussian forms \cite{feynmanpi}, as we discuss in Appendix A. The exact magnitude of $\delta$ will depend on the electrons' separation $d$ and on the interaction time. In this case the wave functions for the $x$ component of the electrons' momentum altered by the interaction become
\begin{eqnarray}
	\label{menos}			\Phi_1^{-}(p) &\equiv& \langle p|\Phi_1^{-}\rangle = \Phi_1(p + \delta),  \\ 
	\label{mais}			\Phi_2^{+}(p) &\equiv& \langle p|\Phi_2^{+}\rangle = \Phi_2(p - \delta), 
\end{eqnarray}
which correspond to momentum shifts of $\mp \delta$ in the wave functions. We note that electron e$_{1}$ gains a negative momentum while electron e$_{2}$ gains a positive momentum of the same amplitude.

It is possible to analyze the quantum states associated with each of the electrons separately by taking the partial traces over the post-selected state of Eq. (\ref{entstate}). In this way, the state $\rho_{1}$ associated to electron e$_{1}$ is
\begin{eqnarray}\nonumber\label{rho1}
\rho_{1} &=& \mathrm{Tr}^{(2)}(|\Phi_{ps}\rangle\langle\Phi_{ps}|)\\\nonumber
         &=& |\Phi_{1}\rangle\langle\Phi_{1}| + Ie^{-i\alpha}\cos(\phi)|\Phi_{1}\rangle\langle\Phi^{-}_{1}|\\
         &&+ Ie^{i\alpha}\cos(\phi)|\Phi^{-}_{1}\rangle\langle\Phi_{1}| + \cos^2(\phi)|\Phi^{-}_{1}\rangle\langle\Phi^{-}_{1}| \label{9} ,
\end{eqnarray}
apart from a normalization factor, with
\begin{equation}\label{I}
I = \int{\Phi_{2}(p)\Phi_{2}(p - \delta)}dp = \exp\left(-\frac{\delta^2}{4W^2}\right). 
\end{equation}
					In the same manner, the state  $\rho_{2}$ associated with the electron e$_{2}$ is
\begin{eqnarray}\nonumber\label{rho2}
\rho_{2} &=& |\Phi_{2}\rangle\langle\Phi_{2}| + Ie^{-i\alpha}\cos(\phi)|\Phi_{2}\rangle\langle\Phi^{+}_{2}| \\
&&+ Ie^{i\alpha}\cos(\phi)|\Phi^{+}_{2}\rangle\langle\Phi_{2}| + \cos^2(\phi)|\Phi^{+}_{2}\rangle\langle\Phi^{+}_{2}|,
\end{eqnarray}
apart from a normalization factor.

Both states $\rho_{1}$, from Eq. (\ref{rho1}), and $\rho_{2}$, from Eq. (\ref{rho2}), which were derived from the entangled pure state of Eq. (\ref{entstate}), represent mixed states for the electrons e$_{1}$ and e$_{2}$ individually. We are able to obtain the probability distributions for the electrons' momenta as $P_{1}(p)=\mathrm{Tr}(\rho_{1}|p\rangle\langle p|)$ and $P_{2}(p)=\mathrm{Tr}(\rho_{2}|p\rangle\langle p|)$, obtaining
\begin{eqnarray}  \label{pp1} \nonumber
 P_{1}(p) &=& \Phi_1^2(p) + \cos^2(\phi)\Phi_1^2(p + \delta)\\
          && + 2I\cos(\phi)\cos(\alpha)\Phi_1(p)\Phi_1(p + \delta) \label{12},
\end{eqnarray}
\begin{eqnarray}
\nonumber P_{2}(p) &=& \Phi_2^2(p) + \cos^2(\phi)\Phi_2^2(p - \delta) \\
\label{pp2} && + 2I\cos(\phi)\cos(\alpha)\Phi_2(p)\Phi_2(p - \delta),
\end{eqnarray}
apart from normalization factors. Both probability distributions have the same form except for a sign change in $\delta$.  

			\begin{figure}
				\centering
					\includegraphics[width=6.0cm]{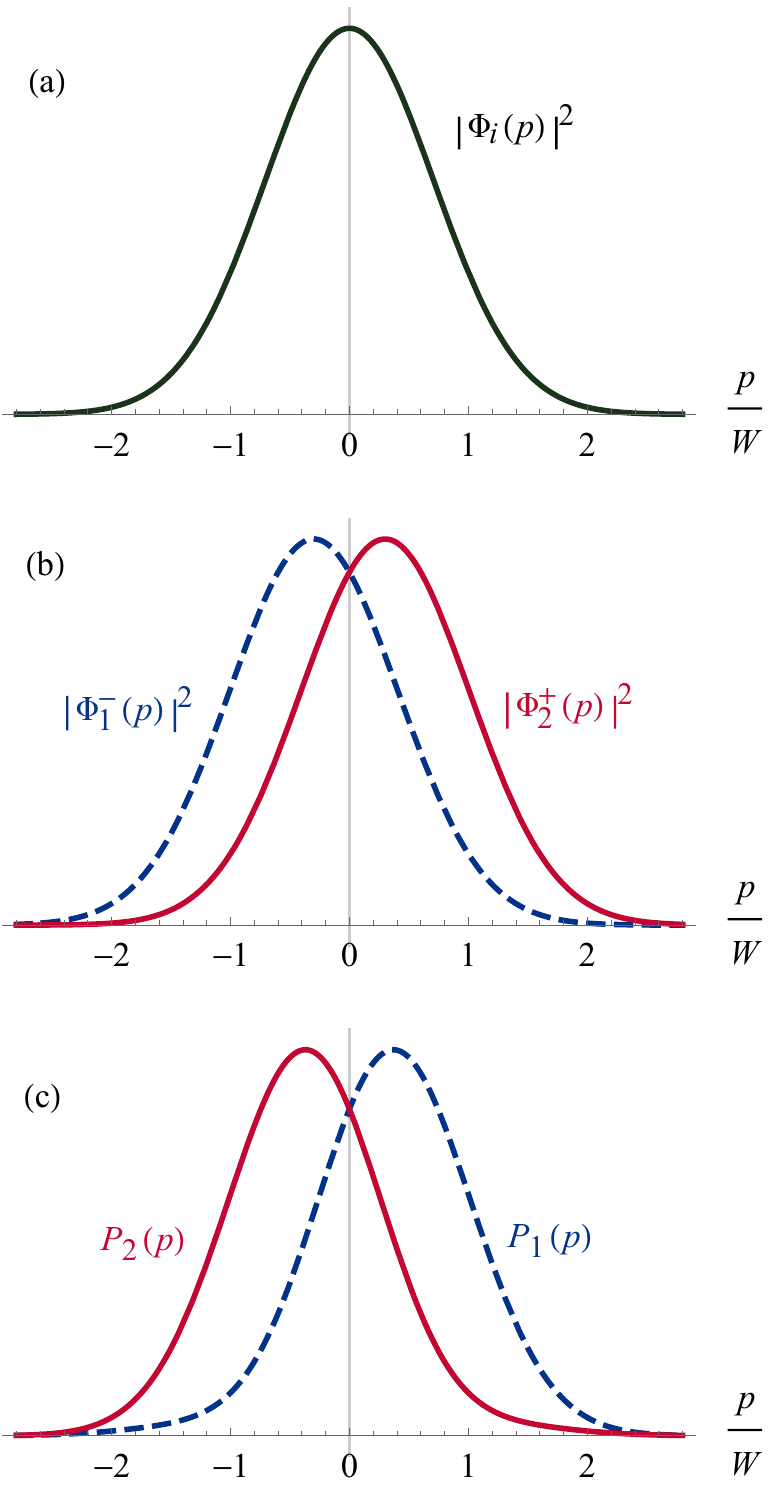}
				\caption{Distributions for the $x$ component of the electrons' momentum wave functions. (a) Initial momentum distribution for each electron, given by $|\Phi_i(p)|^2$ with $\Phi_i(p)$ from Eq. (\ref{gauss}). (b) Momentum distributions of the situations where the electrons propagate through the same path of the interferometer, given by $|\Phi_1^-(p)|^2$ and $|\Phi_2^+(p)|^2$, with $\Phi_1^-(p)$ and $\Phi_2^+(p)$ from Eqs. (\ref{menos}) and (\ref{mais}), for $\delta = 0.3 W$. (c) Momentum distributions corresponding to the quantum superposition of the two situations, given by Eqs. (\ref{pp1}) and (\ref{pp2}) with the parameters $\delta = 0.3 W$, $\phi = {3\pi}/{4}$, and $e^{i\alpha} = 1$. We see that the quantum superposition of an electrostatic repulsion between the electrons with no interaction may result in an effective attraction between them.}
				\label{twoshifts}
			\end{figure}	

Figure \ref{twoshifts} displays the counterintuitive result that we want to emphasize in our paper. Figure \ref{twoshifts}(a) shows the initial distributions of the $x$ component of the electrons' momenta, given by the modulus squared of the momentum wave function of Eq. (\ref{gauss}). Figure \ref{twoshifts}(b) shows the momentum distributions for the situations where the electrons propagate through the same path in the interferometer, given by the modulus squared of the momentum wave functions of Eqs. (\ref{menos}) and (\ref{mais}) with $\delta = 0.3 W$. The momentum distribution for electron e$_1$ is displaced for negative values and the distribution for electron e$_2$ is displaced for positive values, evidencing the repulsive character of the interaction. Figure \ref{twoshifts}(c) shows the momentum distributions predicted by Eqs. (\ref{pp1}) and (\ref{pp2}) with the parameters $\delta = 0.3 W$, $\phi = {3\pi}/{4}$, and $e^{i\alpha} = 1$. The momentum distribution for electron e$_1$ is displaced for positive values and the distribution for electron e$_2$ is displaced for negative values, a result that indicates an effective attractive interaction during their propagation through the interferometer.

\section{Discussion}			
\begin{figure}
				\centering
					\includegraphics[width=8.5 cm]{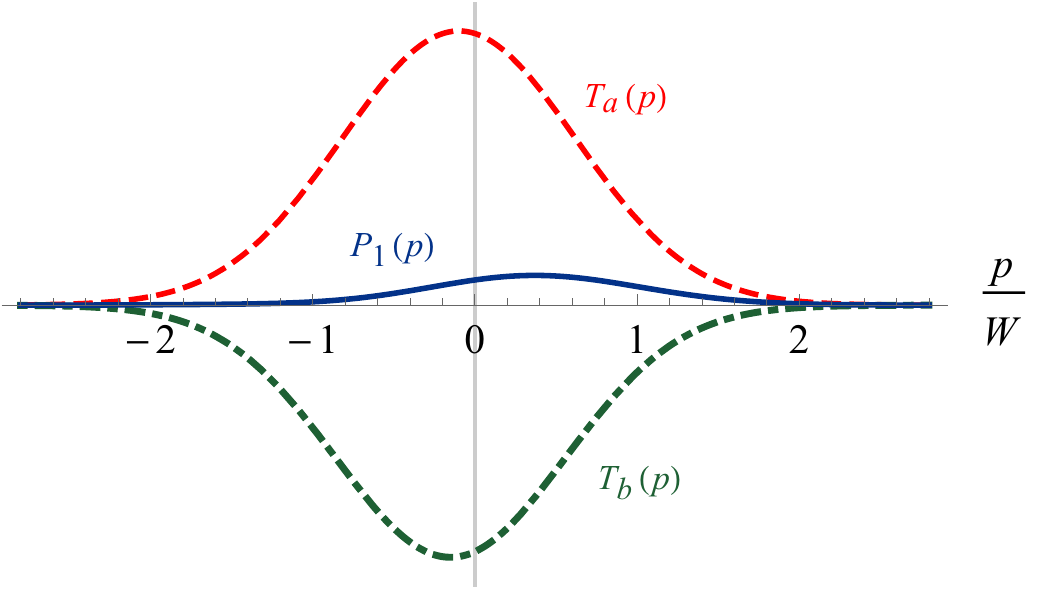}
				\caption{Terms of Eq. (\ref{pp1}). $T_a(p)\equiv\Phi_1^2(p) + {\cos^2(\phi)\Phi_1^2(p + \delta)}$ (dashed red line), $T_b(p)\equiv2I\cos(\phi)\cos(\alpha)\Phi_1(p)\Phi_1(p + \delta)$ (dot-dashed green line) and $P_{1}(p)=T_a(p)+T_b(p)$ (continuous blue line) with the parameters $\delta = 0.3 W$, $\phi = {3\pi}/{4}$, and $e^{i\alpha} = 1$.}
				\label{fig:twopart2}
			\end{figure}

	The strange behavior of an effective electrostatic attraction between electrons in the interferometer is the result of a quantum interference effect. In Fig. \ref{fig:twopart2} we plot the terms of Eq. (\ref{pp1}) that result in the post-selected momentum distribution for electron e$_{1}$ with the same parameters $\delta = 0.3 W$, $\phi = {3\pi}/{4}$, and $e^{i\alpha} = 1$. We note that the term $T_b(p)\equiv 2I\cos(\phi)\cos(\alpha)\Phi_1(p)\Phi_1(p + \delta)$ is the one responsible for the shift to a positive mean value of momentum,  since it subtracts more from the term $T_a(p)\equiv \Phi_1^2(p) + \cos^2(\phi)\Phi_1^2(p + \delta)$ for negative values of $p$ than for positive values of $p$, resulting in a positive average momentum for the distribution  $P_{1}(p)$. The term $T_b(p)$ is the one that comes from the crossed terms, being the result of the interference between the situation where the electrons repel each other with the situation with no interaction.

\begin{figure}
	\centering
		\includegraphics[width=8.0cm]{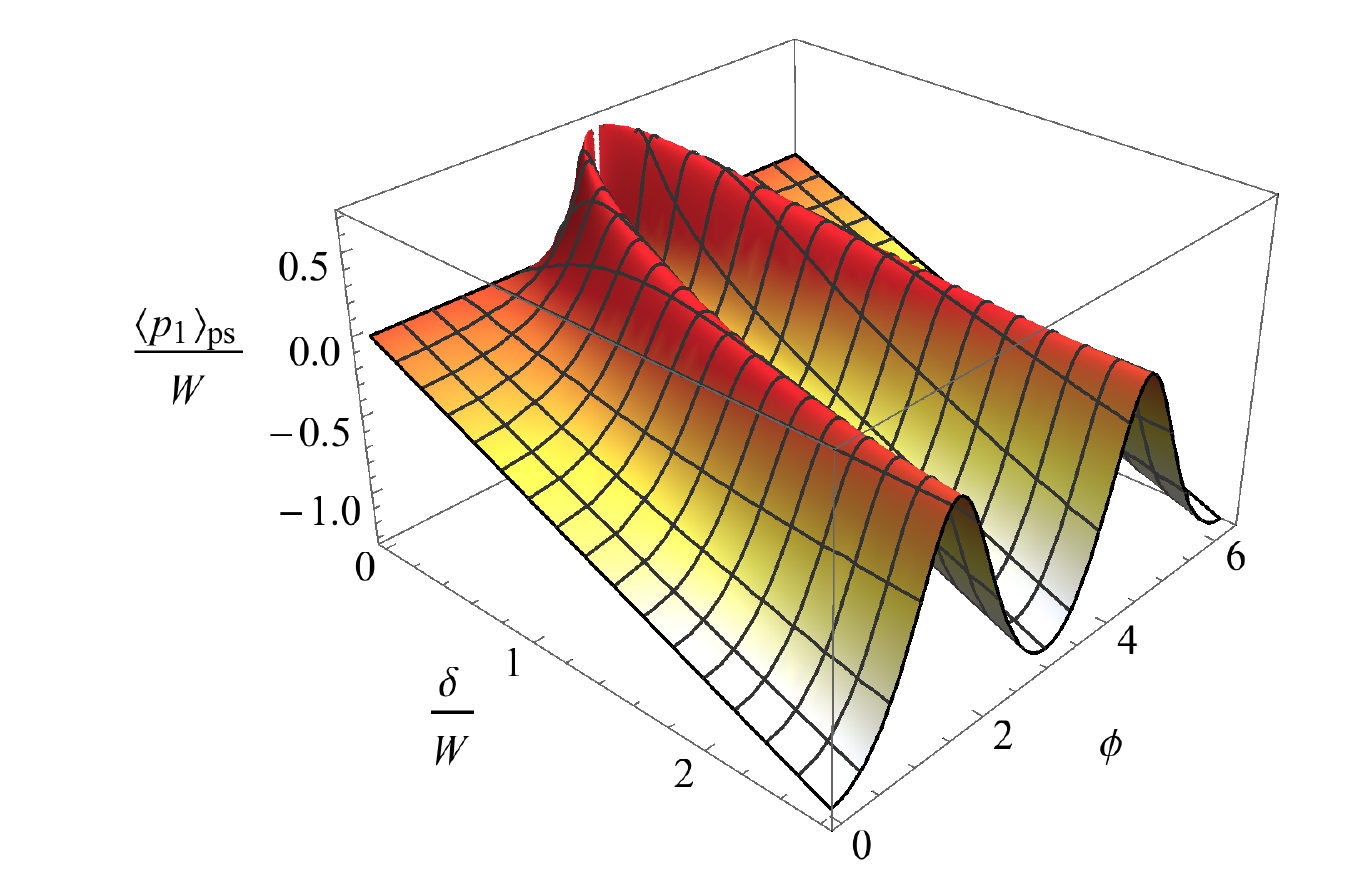}
	\caption{Expectation value of the average momentum of electron e$_1$ normalized by the width of the distribution,  ${\langle p_{1}\rangle_{ps}}/{W}$, as a function of ${\delta}/{W}$ and $\phi$. Anomalous positive values, associated with an effective electrostatic attraction between the electrons, are evident.}
	\label{fig:3d}
\end{figure}

      The expectation value of the momentum of electron 1 leaving the interferometer at the post-selection condition is
\begin{eqnarray}\nonumber
\langle p_{1}\rangle_{ps} &=& \frac{\int_{-\infty}^{\infty}dp\ P_{1}(p)\ p}{\int_{-\infty}^{\infty}dp\ P_{1}(p)}\\
                          &=& \frac{-\delta\left[\cos^2(\phi) + \cos(\phi)\exp\left(\frac{-\delta^2}{4W^2}\right)\right]}{1 + \cos^2(\phi) + 2\cos(\phi)\exp\left(\frac{-\delta^2}{4W^2}\right)} \label{14},
\end{eqnarray}
with $P_{1}(p)$ given by Eq. (\ref{pp1}) with $e^{i\alpha}=1$. It is straightforward to show that $\langle p_{2}\rangle_{ps} = -\langle p_{1}\rangle_{ps}$. The anomalous behavior of an effective attraction between the electrons depicted in Fig. \ref{twoshifts} happens for many values of the interferometer parameters.  In Fig. \ref{fig:3d} we plot the value of ${\langle p_{1}\rangle_{ps}}/{W}$ as a function of the parameters ${\delta}/{W}$ and $\phi$ for $e^{i\alpha}=1$. We note that anomalous positive values for ${\langle p_{1}\rangle_{ps}}/{W}$ occur in a large range of parameters.

It is important to mention that, independently of the parameters used in the interferometer, the average interaction between the electrons is always repulsive when we consider all possible events, without post-selection. This means that if the post-selection of electron e$_1$ exiting by $D_1$ and electron e$_2$ exiting by $C_2$ results in an effective attraction between them, as in the situation depicted in Fig. \ref{twoshifts}, the average interaction in the other situations (electron e$_1$ by $D_1$ and electron e$_2$ by $D_2$, electron e$_1$ by $C_1$ and electron e$_2$ by $C_2$, electron e$_1$ by $C_1$ and electron e$_2$ by $D_2$) is necessarily repulsive, such that the average total interaction is repulsive. We demonstrate this behavior in Appendix B, showing the agreement with the Ehrenfest theorem in this situation, which is a way to say that the average momentum is conserved when one does not post-select the results.

The setup we discuss in this paper is intimately connected to weak measurements \cite{aharonov88,dressel14}. There is a pre- and post-selection of the quantum state of the electrons with the selection of the entrance and exit interferometer ports. The center of the momentum wave function of each electron can be considered a pointer used to measure an observable that indicates if the electrons propagate in the same arm of the interferometer or not. In the weak measurements formalism, the pointer displacement is proportional to the ``weak value'' of the observable, which depends on the pre- and post-selection states. In this situation, it is possible to obtain anomalous weak values for the observable \cite{aharonov88,dressel14}, and the effective electrostatic attraction between the electrons that we study here would be a manifestation of this anomaly.

Now we discuss an experimental proposal for observing the effective electrostatic attraction between electrons due to quantum interference.  Electronic Mach-Zehnder interferometers in free space can be implemented using diffraction gratings acting as mirrors and beam splitters \cite{godun01,gronniger06}. Highly coherent ultrashort electron beams can be generated by laser-triggered emissions from metal tips \cite{hommelhoff06,barwick07,ehberger15}, and it is possible to have the emission of at most one electron per laser pulse \cite{lougovski11}. The optimal coherence properties of such electron beams, as well as their precise time emission with the incidence of a femtosecond laser pulse in the metal tips, could be used to implement the incidence of two electrons at the same time in a Mach-Zehnder interferometer, coming from two tips illuminated by the same laser beam. Consider that the produced electron beams have a transverse width ${\Delta x_{0} \approx 10\ \mu}$m at the entrance of the interferometer, corresponding to a transverse momentum spread $2W \approx \hbar/\Delta x_{0}\approx 10^{-29}\ \T{kg m/s}$, and a kinetic energy around ${10\ \T{ eV}}$, corresponding to a longitudinal velocity $v \approx 2\times 10^{6}$ m/s. If the parallel electron beams are separated by a distance $d\approx 2\ \T{mm}$ and propagate through an interferometer with length ${L\approx4\ \T{cm}}$, the total momentum exchange between the electrons is $\delta = Ft$, where $F={q^2}/({4\pi\epsilon_{0}d^2})$ is the electrostatic force and $t=L/v$ is the interaction duration. The value for $\delta$ for the considered parameters is around 20\% of $W$, ideal for an observation of the effect. For electron emissions that last 100 fs, the initial longitudinal width of the electron wave functions is around $200$ nm for ${v \approx 2\times 10^{6}\ \T{m/s}}$, and increases to around $\approx 10\  \mu$m with the propagation through the interferometer, as shown in Appendix A, being always much smaller than the considered separation $d$ between the electrons. So the components of the forces that act on the electrons on their propagation direction are negligible compared to the transverse forces, which justifies our one-dimensional analysis of the dynamics. More details are presented in Appendix A. A thin metallic foil can be placed between the interferometer arms to avoid the interaction between the electrons when they propagate through opposite paths. Though an optimal technical implementation may be challenging, these considered parameters are  within the scope of what could be experimentally achieved with existing techniques \cite{godun01,gronniger06,hommelhoff06,barwick07,ehberger15}. 
 
To conclude, we have shown that the quantum superposition of the electrostatic repulsion between two electrons (when they propagate in the same arm of an interferometer) with an absence of interaction (when they propagate in opposite arms) may result in an effective electrostatic attraction between them, given the appropriate post-selection. So, in this scenario, the common sense that two charges of equal sign always repel each other is violated due to a quantum interference effect. As we have discussed, an experimental observation of such effect is, in principle, feasible.

This work was supported by the Brazilian agencies CNPq, CAPES, and FAPEMIG.\\

\appendix

\section{Electron propagation through the interferometer arms}

Here we discuss the change of the electrons' wave functions with the propagation through the interferometer arms. In particular, we justify the modification of the $x$ component of the electrons' momentum wave functions from Eq. (\ref{gauss}) to Eqs. (\ref{menos}) and (\ref{mais}) when they propagate in the same arm. The parameters used here are experimentally achievable, as discussed in the end of the main text of the paper.

Let us first consider a free propagation, as when the electrons propagate in opposite arms. Consider that the electrons have a longitudinal velocity $v \approx 2\times 10^{6}$ m/s and the interferometer length is ${L\approx4\ \T{cm}}$, such that the electrons' propagation time is $t\approx 2\times10^{-8}$ s. The width of a Gaussian beam as a function of time can be written as ${\Delta x(t)=\Delta x_0\sqrt{1+\hbar^2t^2/(4m^2\Delta x_0^4)}}$ \cite{cohen}. If the transverse beam waists are ${\Delta x_{0} \approx 10\ \mu}$m at the entrance of the interferometer, the change of the beam widths with propagation for these parameters is of the order of 0.01\% and thus negligible. If the initial beam longitudinal widths are around $200$ nm, the longitudinal spread is considerable, and after a 4 cm propagation the longitudinal widths would be around $6\ \mu$m with the considered parameters, comparable to the considered transverse widths.

Consider that the separation between the electrons in the interferometer is  $d\approx 2\ \T{mm}$, 200 times greater than the dimensions of the electrons' wave functions considered in the previous paragraph. In this case, the electrostatic potential energy when they propagate in the same interferometer arm can be written as 
\begin{eqnarray}\label{electrostatic}\nonumber
	\frac{q^2}{4\pi\varepsilon_0\sqrt{(d+x_2-x_1)^2+(y_2-y_1)^2+(z_2-z_1)^2}}\\
	\approx \frac{q^2}{4\pi\varepsilon_0 d} +  \frac{q^2 x_1}{4\pi\varepsilon_0 d^2} -  \frac{q^2x_2}{4\pi\varepsilon_0 d^2},\ \ \ 
\end{eqnarray}
where $(x_1, y_1, z_1)$ and $(x_2, y_2, z_2)$ represent the electrons' positions in reference frames centered on each beam axis, $q$ is the electron charge, and terms with $x_1x_2/d^3$, $y_1y_2/d^3$, etc. were discarded in relation to terms with $1/d$, $x_1/d^2$, and $x_2/d^2$. The first term on the right side of the above equation is responsible for the phase $\alpha$ from Eqs. (\ref{pa}) and (\ref{pb}). Its presence in the quantum Hamiltonian evolving for a time $t$ results in $\alpha=-q^2t/(4\pi\varepsilon_0\hbar d)$. By varying the distance $d$, $\alpha$ may be adjusted to be an integer multiple of $2\pi$, such that $\rme^{i\alpha}=1$. For the considered parameters, we have $\alpha\approx 7\pi$. However, $d \approx 2.3$mm gives $\alpha\approx 6\pi$.

The system Hamiltonian can  be divided in a Hamiltonian for the $y$ and $z$ components of the electrons' momenta, which generates free propagation and reflections by the interferometer mirrors, and Hamiltonians for the $x$ momentum component of each electron, that includes the electrostatic interaction given by the second or third term on the right side of Eq. (\ref{electrostatic}). For a Gaussian beam with width ${\Delta x_{0} \approx 10\ \mu}$m in the $x$ direction, the maximum momentum component with non-negligible amplitude probability is around $2W \approx \hbar/\Delta x_{0}\approx 10^{-29}\ \T{kg m/s}$, contributing to the kinetic energy with a value $(2W)^2/(2m)\approx 6 \times 10^{-29}$ J, where $m$ is the electron mass. The contribution of the potential energy for the Hamiltonian of each electron is of the order of $q^2 \Delta x_{0}/(4\pi\varepsilon_0 d^2)\approx 6\times10^{-28}$ J, one order of magnitude greater than the kinetic energy term.  

Disregarding the kinetic energy term in relation to the electrostatic interaction term in the Hamiltonians $H_{1}$ and $H_{2}$ that govern the $x$ component of each electron momentum evolution when they propagate in the same arm of the interferometer, the evolution operator for each electron can be written as
\begin{equation}
	U_{i}(t) = \exp\left[\frac{-iH_{i}t}{\hbar}\right]\approx \exp\left[\mp \frac{i\delta x_{i}}{\hbar}\right],
\end{equation}
with $\delta=q^2t/(4\pi\varepsilon_0 d^2)$ and the minus sign referring to electron $e_1$ and the plus sign to electron $e_2$. The above evolution operators are momentum displacement operators, that displace the eigenvalue of a momentum eigenvector by amounts $\pm\delta$. So the application of the above evolution operators in states described by the momentum wave function of Eq. (\ref{gauss}) results in states with momentum wave functions given by Eqs. (\ref{menos}) and (\ref{mais}). For the considered parameters, we have $\delta\approx 10^{-30}\ \T{kg m/s}\approx W/5$, which is a reasonable value for observing the quantum effects we discuss in this work. 

The electrostatic repulsion between the electrons generates a beam tilt on the phase fronts perpendicular to the propagation direction for each electron wave function, and one may wonder if the interference of this probability amplitude with the probability amplitude with no interaction and no beam tilt could result in an extra effect of interference fringes, unconsidered in this work. But note that if we interfere a beam with zero average $x$ component of momentum with a beam displaced in momentum by $\delta \approx 10^{-30}\ \T{kg m/s}$ in the $x$ direction, this results in interference fringes with spacing around $h/\delta\approx6\times10^{-4}$ m at the interferometer exit. Since the considered beam diameter is around $10^{-5}$ m, a much smaller value, there will be no interference fringes in this condition.

\section{Agreement with the Ehrenfest theorem}

As we have mentioned, although we are able to observe the anomalous effect of attraction between two electrons in the system when the appropriate post-selection of exit ports is made, it is necessary that the average interaction between the electrons be repulsive overall. This expectation is derived from the Ehrenfest theorem, which states that the behavior of the averages of quantum observables should agree with those expected classically. Here we show how the Ehrenfest theorem applies to our interferometer.

First we note that there are in total $4$ possibilities of paths for the two particles inside the apparatus, and $4$ possible ways that they can leave the system at the end of the experiment, making up for a total of $16$ evolution possibilities for the system. This means that a priori we have a $16$ term superposition for our complete two-electron state after they leave the apparatus. The four-term superposition for the joint state just before the action of BS$_{2}$ can be written as: 
\begin{widetext}
\begin{equation} \label{bbs2}
 irte^{2i\phi}e^{i\alpha}|\Phi^{-}_{1},A_{1}\rangle|\Phi^{+}_{2},A_{2}\rangle +irte^{i\alpha}|\Phi^{-}_{1},B_{1}\rangle|\Phi^{+}_{2},B_{2}\rangle  +t^2e^{i\phi}|\Phi_{1},A_{1}\rangle|\Phi_{2},B_{2}\rangle-r^2e^{i\phi}|\Phi_{1},B_{1}\rangle|\Phi_{2},A_{2}\rangle, 
\end{equation}
where we have taken into account the existence or not of an interaction between the electrons and the appropriate phase gains due to the evolution of the system as done in our previous discussion. 

The effect of BS$_{2}$ over the different terms of this state can be written as
\begin{eqnarray}\label{basis} 
\nonumber |A_{1}\rangle|A_{2}\rangle &\Rightarrow& t^2|C_{1}\rangle|C_{2}\rangle - r^2|D_{1}\rangle|D_{2}\rangle  +irt(|C_{1}\rangle|D_{2}\rangle + |D_{1}\rangle|C_{2}\rangle) ,\\
\nonumber |B_{1}\rangle|B_{2}\rangle &\Rightarrow& -r^2|C_{1}\rangle|C_{2}\rangle + t^2|D_{1}\rangle|D_{2}\rangle +irt(|C_{1}\rangle|D_{2}\rangle + |D_{1}\rangle|C_{2}\rangle) ,\\
\nonumber |A_{1}\rangle|B_{2}\rangle &\Rightarrow& irt(|C_{1}\rangle|C_{2}\rangle + |D_{1}\rangle|D_{2}\rangle)  +t^2|C_{1}\rangle|D_{2}\rangle -r^2|D_{1}\rangle|C_{2}\rangle , \\
 |B_{1}\rangle|A_{2}\rangle &\Rightarrow& irt(|C_{1}\rangle|C_{2}\rangle + |D_{1}\rangle|D_{2}\rangle) -r^2|C_{1}\rangle|D_{2}\rangle + t^2|D_{1}\rangle|C_{2}\rangle . 
\end{eqnarray}

  By plugging Eqs. (\ref{basis}) into Eq. (\ref{bbs2}), we reach the final joint state superposition for the electrons leaving the system:  
	\begin{eqnarray}	
\nonumber  irt\Big[e^{i\alpha}(t^2e^{i\phi}-r^2e^{-i\phi})|\Phi^{-}_{1}\rangle|\Phi^{+}_{2}\rangle +(t^2-r^2)|\Phi_{1}\rangle|\Phi_{2}\rangle\Big]&&|C_{1}\rangle|C_{2}\rangle\\
\nonumber	+\ irt\Big[e^{i\alpha}(t^2e^{-i\phi}-r^2e^{i\phi})|\Phi^{-}_{1}\rangle|\Phi^{+}_{2}\rangle +(t^2-r^2)|\Phi_{1}\rangle|\Phi_{2}\rangle\Big]&&|D_{1}\rangle|D_{2}\rangle\\
\nonumber	+\ \Big[-r^2t^2e^{i\alpha}(e^{i\phi}+e^{-i\phi})|\Phi^{-}_{1}\rangle|\Phi^{+}_{2}\rangle +(t^4 + r^4)|\Phi_{1}\rangle|\Phi_{2}\rangle \Big] &&|C_{1}\rangle|D_{2}\rangle\\
 -\ r^2t^2\Big[e^{i\alpha}(e^{i\phi}+e^{-i\phi})|\Phi^{-}_{1}\rangle|\Phi^{+}_{2}\rangle +2|\Phi_{1}\rangle|\Phi_{2}\rangle\Big] &&|D_{1}\rangle|C_{2}\rangle \label{phiT}.	
	\end{eqnarray}
	\end{widetext}
	The post-selection of exit ports made previously in our discussion meant projecting this state superposition in the vector state $|D_{1}\rangle|C_{2}\rangle$, and by doing this we get the wave function of Eq. (\ref{entstate}) used to derive our results, as we should.
	
	To show that the average interaction is always repulsive when no post-selection is made, we can focus on what happens to electron e$_{1}$. We have created a situation where electron e$_{1}$ has counterintuitively gained positive momentum due to its interaction with electron e$_{2}$ by post-selecting the exit ports, therefore mimicking an attractive interaction. We can nevertheless show that, on average, the momentum gained by this electron when we consider the complete joint state of Eq. (\ref{phiT}) is always either null or negative. This means that the expectation value of the electron e$_{1}$ momentum without post-selection must be always null or negative, namely
	\begin{eqnarray}\nonumber
	\langle p_{1}\rangle =\ &&\langle p_{1}\rangle_{CC}P_{CC} +	\langle p_{1}\rangle_{DD}P_{DD} + \langle p_{1}\rangle_{CD}P_{CD}\\
	                       &&+ \langle p_{1}\rangle_{DC}P_{DC},
	\end{eqnarray}
where $P_{jk}$ is the probability of detecting electron e$_1$ at exit $j$ and e$_2$ at $k$, and $\langle p_1\rangle_{jk}$ is the respective average momentum for this detection. This quantity can be derived by repeating the process done in Eqs. (\ref{9}), (\ref{12}), and (\ref{14}) for each of the four exit port possibilities, $|C_1\rangle|C_2\rangle$, $|D_1\rangle|D_2\rangle$, $|C_1\rangle|D_2\rangle$, and $|D_1\rangle|C_2\rangle$. Some straightforward algebra shows us that the total average momentum gained by electron e$_{1}$ is
\begin{align}
 \langle p_{1}\rangle &= (t^4+r^4)\langle \Phi_1|p_1|\Phi_1\rangle+2t^2r^2\langle \Phi^-_1|p_1|\Phi^-_1\rangle\notag\\
 &=-2t^2r^2\delta.
\end{align}
This perfectly agrees with our classical intuition, as the first term incorporates the probability that the electrons are either both transmitted or both reflected by BS$_1$ (they do not interact), and the second term considers the probability that one of the electrons is transmitted and the other is reflected at BS$_1$ (they do interact). So the final average momentum is simply the momentum gained when they do interact, $\langle \Phi^-_1|p_1|\Phi^-_1\rangle = -\delta$,  which comes from a repulsive interaction, times the probability of interacting, $2t^2r^2$. Therefore the average interaction is repulsive, in agreement with momentum conservation and the Ehrenfest theorem.

%

\end{document}